# Solid core dipoles and switching power supplies: lower cost light sources?

Jay Benesch, Sarin Philip

**Abstract**

As a result of improvments in power semiconductors, moderate frequency switching supplies can now provide the hundreds of amps typically required by accelerators with zero-to-peak noise in the kHz region ~0.06% in current or voltage mode. Modeling was undertaken using a finite electromagnetic program to determine if eddy currents induced in the solid steel of CEBAF magnets and small supplemental additions would bring the error fields down to the 5ppm level needed for beam quality. The expected maximum field of the magnet under consideration is 0.85T and the DC current required to produce that field is used in the calculations. An additional 0.1% current ripple is added to the DC current at discrete frequencies 360 Hz, 720 Hz or 7200 Hz. Over the region of the pole within 0.5% of the central integrated BdL the resulting AC field changes can be reduced to less than 1% of the 0.1% input ripple for all frequencies, and a sixth of that at 7200 Hz. Doubling the current, providing 1.5T central field, yielded the same fractional reduction in ripple at the beam for the cases checked. For light sources with aluminum vacuum vessels and full energy linac injection, the combination of solid core dipoles and switching power supplies may result in significant cost savings.

**Finite element modeling**

The CEBAF accelerator is a CW electron machine using solid core dipoles throughout. There are a pair of superconducting linacs through which the beam passes multiple times, recirculated by ten arc with different fields. Pathlength varies among arcs so three-magnet chicanes in each arc are used to ensure that the electrons arrive at the next linac on crest, maximizing energy gain and minimizing energy width. It is essential to minimize the pathlength variations due to time dependent field variations driven by power supply ripple. We performed a field modeling study to quantify the required power supply specifications. The smallest magnet among those used in the pathlength-adjusting chicanes was chosen to reduce the clock time required by the modeling. The fraction of the magnet modeled is shown in figure 1. Total steel length is 46cm and coil length is 60cm. The full coil set and a fourth of the steel are shown. This is the worst case among the magnets at CEBAF because the ratio of steel to coil length is smallest. A simulation with central field 0.85T was chosen, one with the peak current required for planned operation in CEBAF. This ensures that the relative permeability throughout is at the minimum expected during operation. Relative permeability is still a few thousand except at core edges. If one is driving magnets into full saturation the response will change; 1.65T is not sufficient to change the result with the experimentally verified BH curve used.

Eddy current shielding is rarely considered in magnet design. Lamination thickness is chosen in magnets for fast-cycling machines so the AC field penetrates the iron completely in a short time. Switching power supplies at 1 kHz have caused issues at JPARC (1). For CW machines and light sources with full-energy linacs, solid core magnets are feasible. Eddy currents will be induced in the steel by current ripple and the fields induced will oppose the external fields. The first author encountered eddy current effects some three decades ago when the aluminum alloy of a thermal shield in an MRI magnet was changed, lowering the net effect of the pulsed gradient coils in the patient volume.

Finite electromagnetic (FEM) software from Vector Fields (VF) was used in this work, the transient solver ELEKTRA/TR. Software with the same capability is available at least from ANSYS (Maxwell) and Integrated Engineering Software (Faraday). All of these allow one to impose an AC current with one or more frequencies in addition to a large DC current which puts the magnetic material in the model at the appropriate location on the BH curve. The VF model preparation software allows one to constrain the mesh via layers parallel to the surface, allowing small dimension in the direction crucial to skin depth modeling and larger in the two transverse dimensions. In calculating skin depth the average relative permeability in the magnet from a DC model, 4000, was used. Ten 110μm-thick layers were used on the surfaces facing the coil and on the pole in the 360 Hz (~187 μm skin depth) cases. At 720 Hz (~133 μm skin depth) models, ten layers at 70μm were used . At 7200 Hz, 20μm layers (~42 μm skin depth). For copper multiply the skin depths by 20; for aluminum 27. The 0.1% ripple term was set by the need to see a ppm or greater change in the field integral along the center of the dipole. One vendor's current and voltage noise specifications (4) are ~0.06% of full scale zero to peak, so 0.1% is a realistic value. All of the software packages cited allow one to couple the coils to an external circuit with engineering parameters. This capability was not used in this exercise. The coils are simply a block with applied current density $J = J_0*(1+0.001\sin(2*pi*f))$.

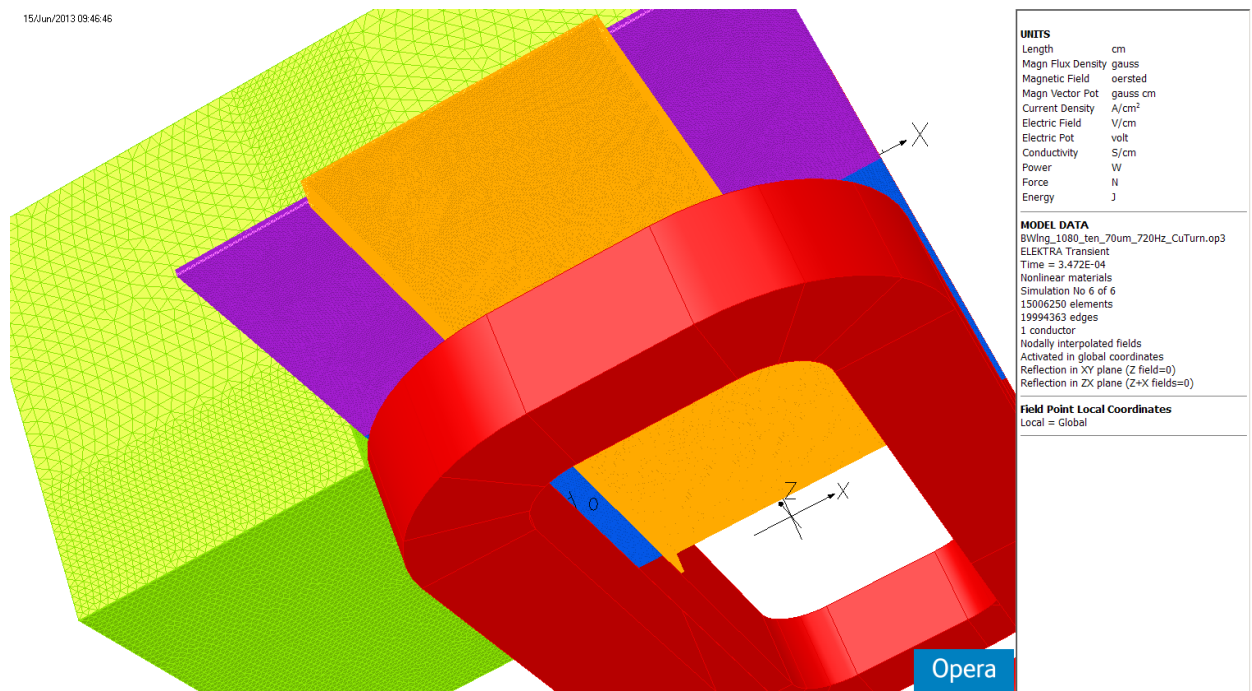

Figure 1. Dipole model with additional eddy current shield material (blue and purple) and vacuum vessel (orange). Pole half-gap is 1.27 cm and plate also starts at y=1.27 cm. Plate thickness 0.3cm, width 24 cm, length 17 cm. Vacuum vessel thickness 0.15 cm, extends 1.5cm beyond plate.

In any FEM problem there is a balance which must be struck among mesh design, convergence parameters, and clock time available. The layering choices mentioned above were obtained by looking at solutions with various meshes beginning at 1mm and working down from there. The basic magnet model was used in arriving at these choices. After development, the vacuum vessel (orange in figure 1), supplemental end plate (purple) and the plates which convert the end plate into a shorted turn (blue) were included in the meshed model for all cases. Only the material conductivities were changed among the six cases presented below. The accuracy of absolute results is determined by the BH curve. Many more points than are available in any BH curve found in the literature are needed to obtain absolute results in agreement with measurement at the part per thousand level. Only relative effects are reported here.

The six cases presented are:
- Basic magnet: coils and bulk steel only. All materials except steel set to zero conductivity
- Vacuum vessel set to 1.45E6 S/m, the conductivity of stainless steel, denoted stainless
- Vacuum vesset set to 3.3E7 S/m, the conductivity of aluminum, denoted aluminum
- End plate (purple) set to 6E7 S/m, the conductivity of copper, denoted CuPlate. Includes stainless vacuum vessel.
- End plate (purple) and coil pocket plates (blue) set to 6E7. Software allowed induced currents to flow across material boundaries if the conductivities are equal, so this produces a shorted copper turn and is denoted CuTurn. Vacuum vessel set to stainless conductivity. (3)
- As above, but vacuum vessel set to aluminum conductivity. Denoted CuTurnAl.

In order to better approximate the case for light sources, the last two cases were also run with 1.64T field: the coil cross section was doubled and the current density increased by 25%. Amp-turns were increased by 2.5 while field increased only by a factor of 1.93, so the steel is well into saturation for these. Meshes were changed because the average relative permeability decreased from ~3300 to ~600 in the steel, increasing skin depth by ~2.3. The lower frequency cases were run with 250 μm layers, 7200 Hz with 70 μm layers.

Aluminum and steel end plates (purple) were also examined. A thicker aluminum plate works as well as the 3 mm copper if space permits. Steel works surprisingly well given that the plate is saturated directly under the coil even in the lower field case. Computational time is much increased because the steel plate response is so non-linear with location and numerical work was on it was not continued. The case denoted CuTurn is a great improvement over reference (2).

The results presented below are field integrals at y=0 with z=[0,40] cm at x=0 and 4 cm. The pole is 6.35 cm half-width and 22.95 cm half-length. Pole face is at y=1.27 cm. The end plate (purple) encompasses x=[-12,12] y=[1.27,1.57] z=[23,39] cm. The 0.05 cm gap between end plate and pole face is due to the software need to keep non-zero conductivities separated in space. The vacuum vessel is 0.15 cm thick with top 0.07cm from the pole face for the same reason. It is 40.5 cm half-length. These lengths, 39 and 40.5 cm, were chosen because they are physically feasible with the existing magnets. The plates cannot be longer without interfering with the vacuum vessel flange bolts. The face of the existing vacuum vessel flange is at z=40.5 cm in this coordinate system. The field integrals are taken only to z=40cm because the dipole field changes sign at z=32.5 cm where the field lines reconnect around the full magnet there. The AC field component beyond this z modestly balances that at smaller z. As will be seen, the cases with aluminum vacuum vessel show such good AC shielding that the small positive field at large z would dominate the results if the integrals were extended out to z=60 cm, where the relatively fine mesh ends. Coarse mesh extends a few meters more so the field at the edge of the model is under 0.2 G. The first author chose not to extend the vacuum vessel this far to keep the model size tractable; it's still 15M elements.

In Figure 2 we plot the model results for the basic model, steel and coil only, at the three frequencies and two locations, the pole midline x=0 and the closest approach to the pole edge in the dogleg magnet system, x=4 cm in this model. The following graphs show results grouped by frequency at x=4 cm because the full excursion will determine the utility of the scheme.

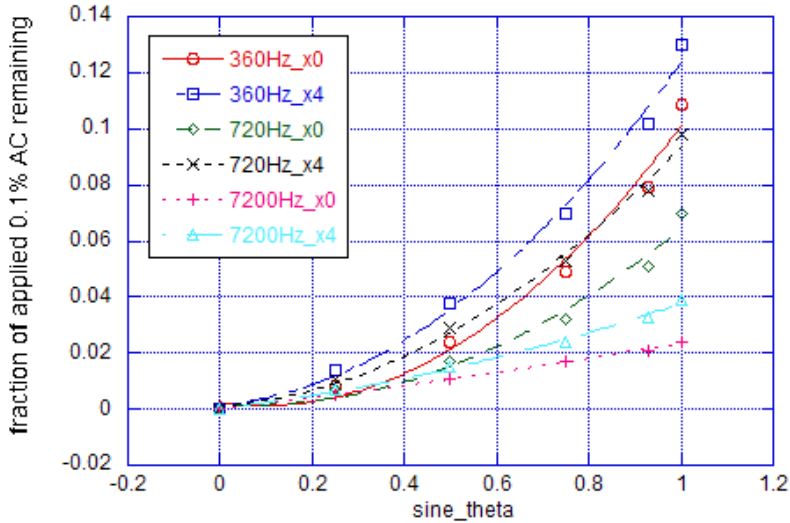

Figure 2. Fraction of applied 0.1% remaining in the 40 cm line integrals along x=0 and x=4 cm for the basic magnet, steel/coil/air only, at 360 Hz, 720 Hz, and 7200 Hz. Quadratic polynomial fits included to guide the eye. Sine = 0.93 point is calculated by the software due to the time difference between the adjacent points; the software is not programmed to recognize the functional form of the drive in defining time steps.

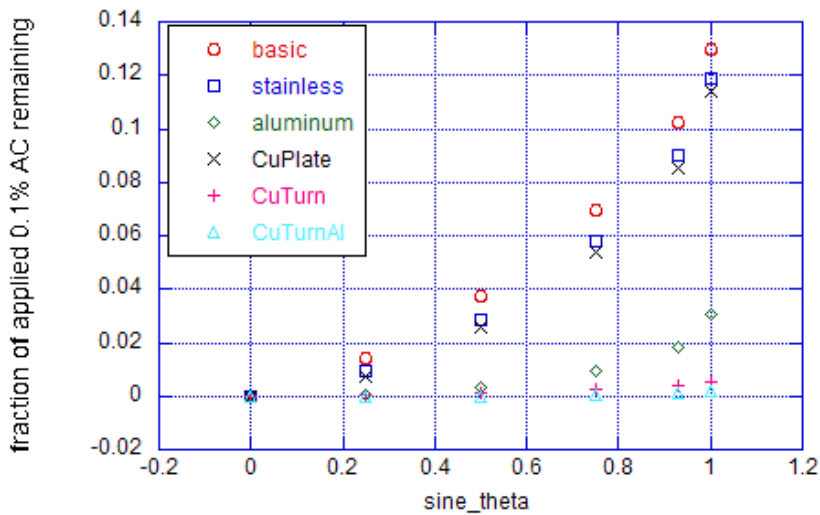

Figure 3. Fraction of applied 0.1% 360 Hz AC component remaining in the line integrals along x=4 cm for the six magnet configurations modeled. The three cases with high-conductivity materials which comprise complete circuits, the shorted copper turn (CuTurn), the aluminum vacuum vessel (aluminum) and the case with both (CuTurnAl) provide much better damping than do the pole steel alone (basic), the low conductivity stainless vacuum vessel, or the supplemental plate which shadows the vacuum vessel beyond the pole (CuPlate). At this frequency the addition of the plate to the stainless vacuum vessel has no useful effect.

The response to 720 Hz AC is very similar. It is shown in figure 4 because for some applications the factor of ten reduction in applied AC current may suffice when coupled with additional reduction provided by circuit elements which are not included in the finite element modeling, for instance capacitors placed across the magnet terminals. The capacitors may be unacceptably large at this frequency in many but not all accelerator applications.

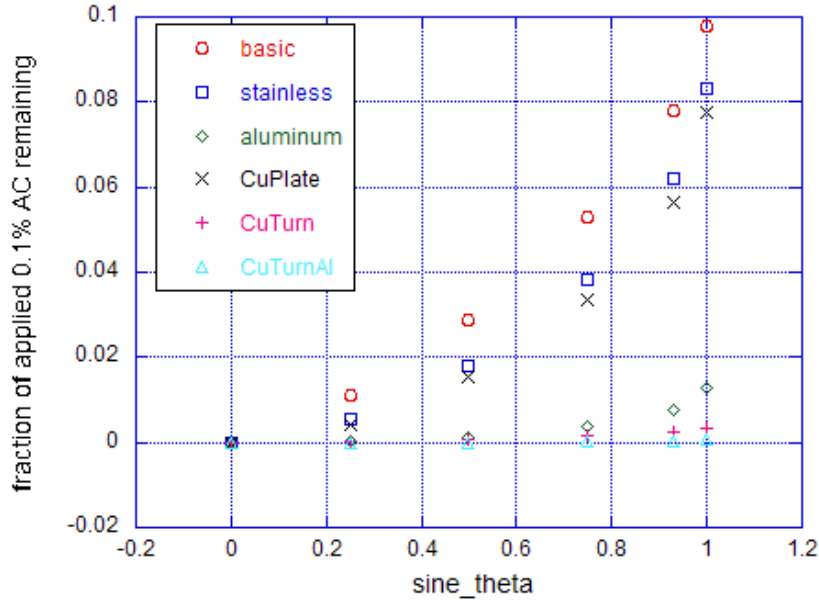

Figure 4. Fraction of applied 0.1% 720 Hz AC component remaining in the line integrals along x=4 cm for the six magnet configurations modeled. Again the configurations with complete circuits at high conductivity are superior.

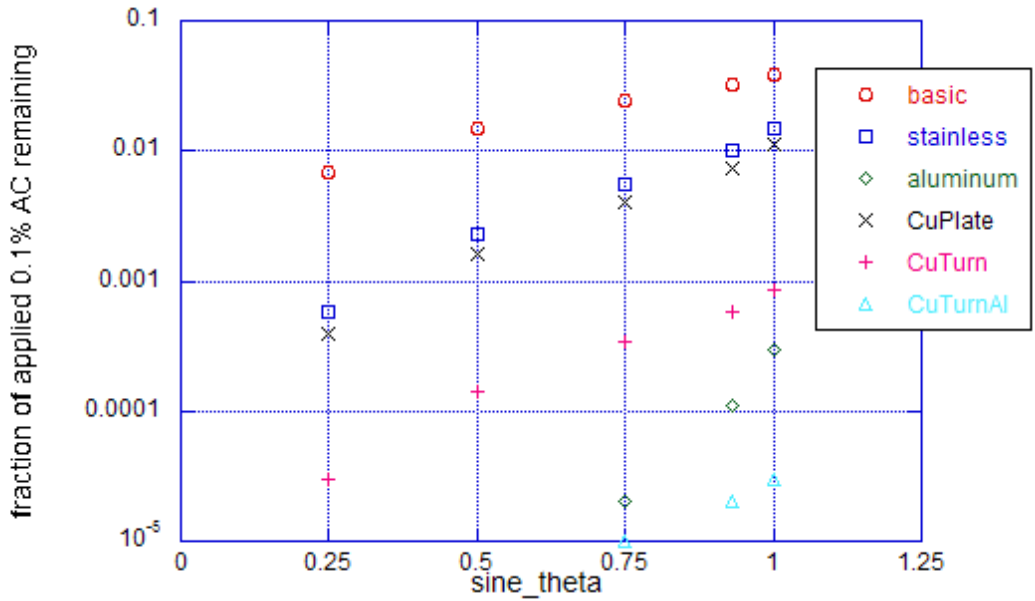

Figure 5. Fraction of applied 0.1% 7200 Hz AC component remaining in the line integrals along x=4 cm for the six magnet configurations modeled. Vertical scale is logarithmic. Values which are below 10ppm on vertical axis suppressed for the two cases with aluminum vacuum

vessel.  The shorted 3 mm thick copper turn with stainless steel vacuum vessel performs well enough for the need at CEBAF, reducing the peak AC component to 1ppm of the DC.  For light sources with obligate aluminum vacuum vessels, these may suffice if 1.5mm thick.  If vessel is thinner or 0.3ppm is not sufficiently damped, add the shorted copper turn.

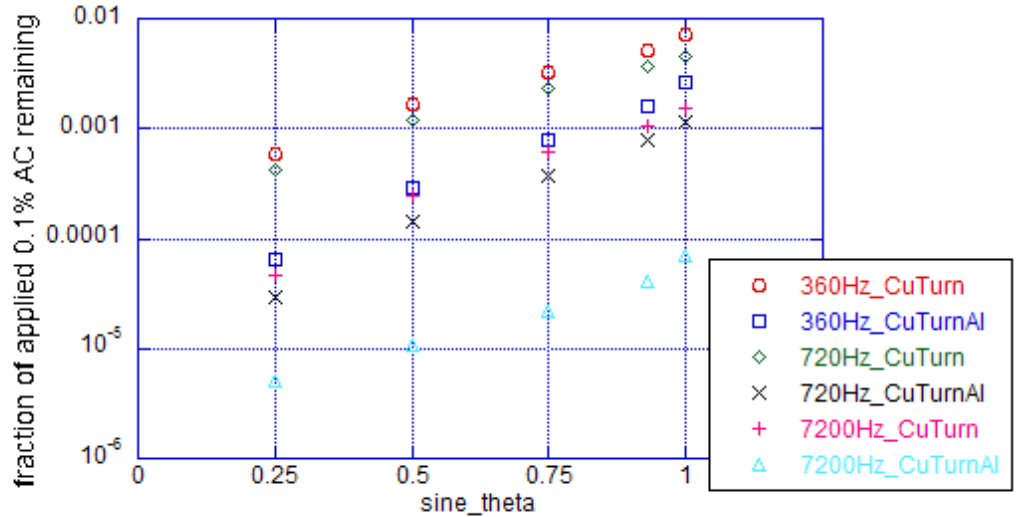

Figure 6.  Fraction of applied 0.1% AC remaining for two magnet cases at three frequencies each.  Amp-turns for these models are 2.5x those in previous figures, yielding 1.64T DC field and 1.93x the field integral.  Cases at 360 Hz and 720 Hz were run with ten 250 micron layers due to increased skin depth (lower relative permeability).  Cases at 7200 Hz were run with ten 70 micron layers.

The effect of the AC component on multipoles was checked on two models, the 0.85T model with 360 Hz applied to solid core dipole with stainless steel vacuum vessel and the 1.6T model with stainless vessel and copper turn at 360 Hz presented in figure 6.  An orbit entering at three degrees with six degrees total bend was used.  Under the poles, this orbit is within x=[-3.4,-4] so the DC multipole content is high.  Within the same Z extent used in the figures, the multipole content is damped at about the same level as the dipole.  Since the multipoles start as a few ppt of the dipole for this extreme orbit, the changes are 21 ppm of dipole for quadrupole, 3 ppm for sextupole, and under 60 ppb for higher multipoles for the 0.85T case.  For the case with shorted copper turn at higher field, all of the changes are under 1.6 ppm of the dipole. (5)

This finite electromagnetic analysis shows that the combination of solid core dipoles and high current switching supplies with frequencies at least 7200 Hz (8-30 kHz typical) are compatible with the requirements for the CEBAF doglegs and will likely limit closed orbit deviations in light source storage rings sufficiently given their obligate aluminum vacuum vessels.  If another order of magnitude in damping is required, adding a shorted turn of 3 mm thick copper under the coil suffices.  As the next section shows, one to two more orders of magnitude in noise reduction may be available by tuning the drive circuit, including placing modest capacitors across the leads of each magnet in the accelerator enclosure.  This will bring the closed orbit deviation in storage rings due to switching supplies to the ppb level.

Circuit model and measurements

A circuit model was created, figure 7. Measured impedances as a function of frequency were used so the effects discussed for the finite element work above are included in the circuit model. The resulting impedance is shown in figure 8. Figure 9 shows that the peak in impedance ~3 kHz is present in real strings of magnets. Finally, the table compares measurements made with a dogleg magnet on the bench at low excitation with the predictions of the model.

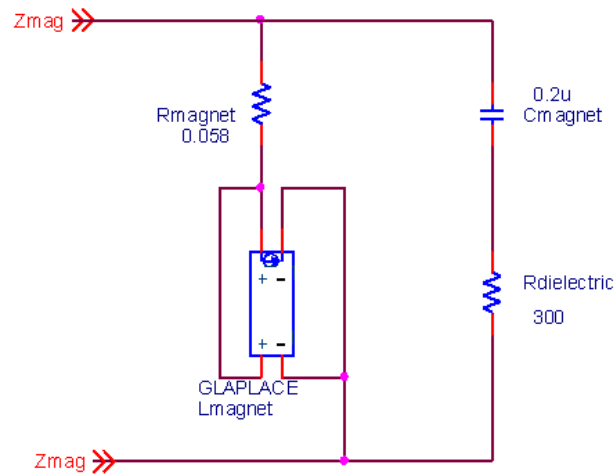

Figure 7. Electrical model for solid core magnet. Inductance L(s) with s complex.

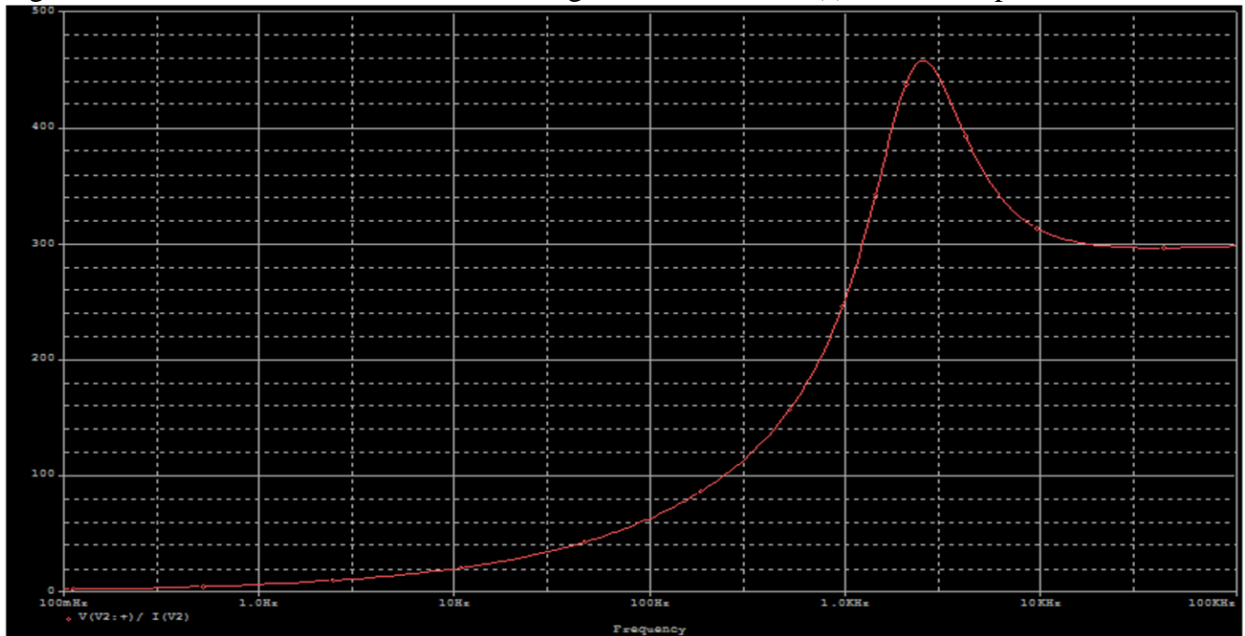

Figure 8. Simulation of load impedance characteristic of magnet

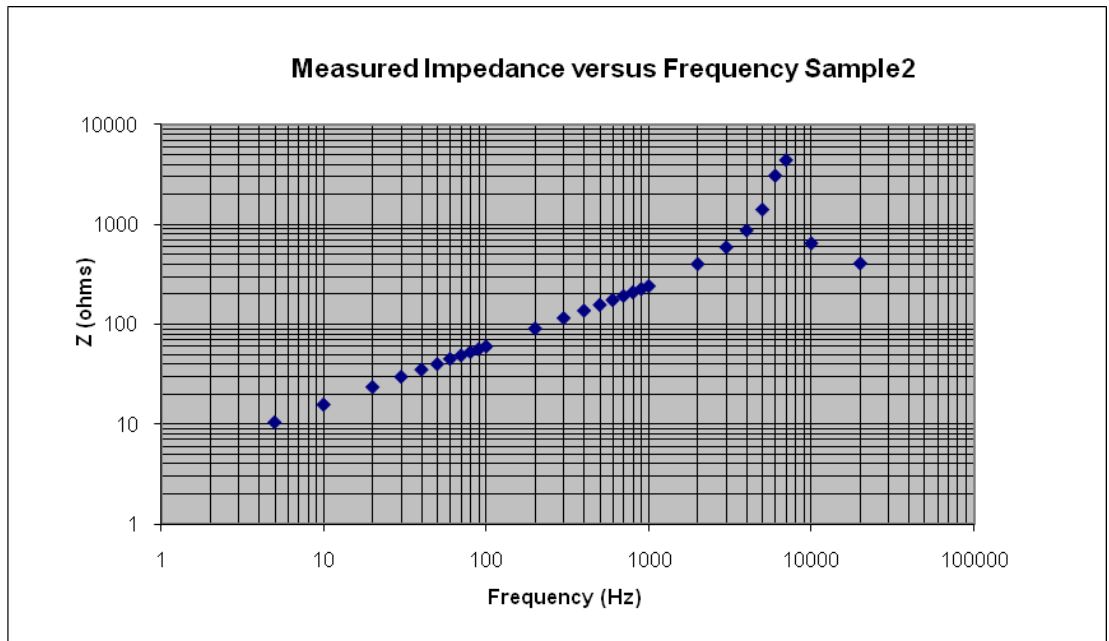

Figure 9. Typical impedance profile of a magnet string, not the single magnet under test. The behavior is similar to that in figure 8, the simulation, although the peak in impedance is at a frequency about a factor of two higher.

| F(Hz) | V rms | I rms (A) | Z ohms | Measured Zac/Zdc | Simulation Zac/Zdc |
|-------|-------|-----------|--------|------------------|---------------------|
| 0     | 0.048 | 1         | 0.048  | 1                | 1                   |
| 9     | 0.1   | 0.146     | 0.685  | 14.3             | 11                  |
| 100   | 0.1   | 0.066     | 1.52   | 31.6             | 30                  |
| 1000  | 0.1   | 0.019     | 5.27   | 110              | 100                 |
| 7000  | 0.1   | 0.009     | 11.1   | 231              | 260                 |

**Table 1.** Comparison of circuit simulation and measured impedances.

These impedance measurements were supplemented with magnetic measurements with a simple pick-up loop of width about 2 cm and about half the length of the magnet. It shows that the AC flux is concentrated at the edges of the pole, in agreement with the finite element model.

Small magnet tests

A 102 mm long, high uniformity magnet was needed for a low energy nuclear physics experiment to be performed in the CEBAF injector (6). Momentum resolution at the 0.01% level is desired by the collaboration. Since the magnet is powered by one of the CEBAF trim power supplies, a regulator card in a rack powered by a pair of three-phase input DC supplies, a 3 mm copper eddy current shield was included in the design, figure 10. Opportunistic magnetic measurements were made with 1% AC at frequencies up to 720 Hz applied on top of DC currents sufficient to fully magnetize the steel but not take it into saturation. The eddy current shield may be made into a cooling plate by soldering a water tube to it; this has not been done so DC currents were limited to the linear regime of the steel. The dB/dt probe used is ~15 x 1.3 $cm^2$ with 150 turns, ~2900 cm$^2$ . The results are shown in Table 2. The response of this magnet to AC has not been modeled. The measurements were made to obtain a qualitative confirmation of the modeling done on the larger magnet discussed above.

**Table II** - Volts seen by probe for excitation 10A*(1+0.01sin(2*pi*f)). Note that Model-CuTurn includes a stainless steel vacuum vessel which was not present during the measurements in the small dipole reported in the table. "Free space V" is the voltage that would be seen by the probe were the magnet laminated yet full density, i.e. with B as with steel but no damping.

| Frequency | Free space V | Bare magnet V | Model | Copper turn V | Model-CuT |
|---|---|---|---|---|---|
| 60 | 0.15 | 0.0186 (12.4%) | | 0.0045 (3%) | |
| 120 | 0.3 | 0.033 (10.8%) | | 0.0047 (1.6%) | |
| 360 | 0.9 | 0.075 (8.3%) | 13% | 0.0052 (0.58%) | 0.54% |
| 720 | 1.8 | 0.111 (6.2%) | 9.8% | 0.006 (0.35%) | 0.32% |

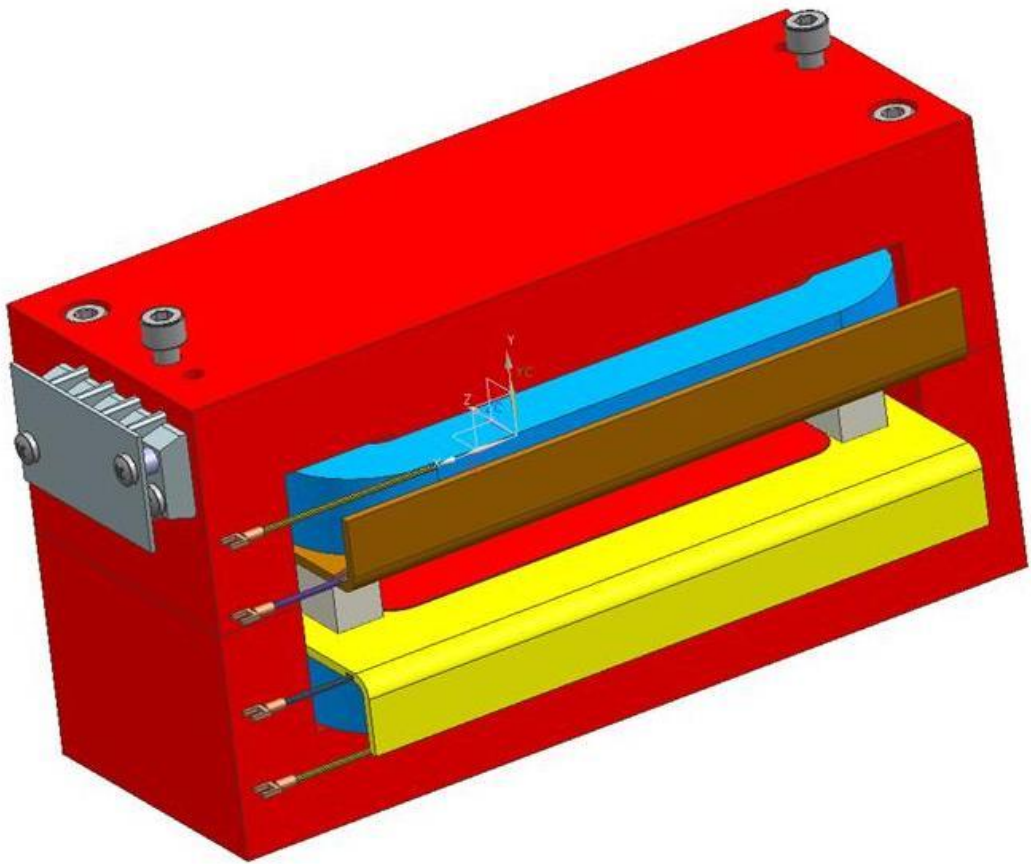

Figure 10. Magnet tested. Steel is red, coils are blue, copper shield plates are brown and yellow. Grey bars between plates are aluminum blocks which secure coils on steel core; they are insulated from the copper by 3mm rubber. Pole length 102 mm, width 160 mm.

Conclusion

We have studied the effect of 0.1 % zero-to-peak AC noise at the terminals of a high current switching power supply on the fields seen by an accelerator beam. Eddy current damping in a solid core dipole and associated stainless steel vacuum vessel reduce the effect two orders of magnitude. Aluminum vacuum vessel or a 3 mm thick shorted copper turn under the coil provide another one to two orders of magnitude. Drive circuit tuning, including distributed capacitors, can lower the noise seen by the beam to the part per billion level, far lower than the noise specification of high current linear supplies. Magnetic measurements on a small magnet with 3 mm shorted copper turn show comparable reduction to the finite element model. It follows that light sources with full energy linac injectors may construct the storage ring at substantially lower cost by specifying solid core magnets and high current switching supplies in place of laminated magnets and high precision linear supplies.

References

1. Y. Watanabe et al., IPAC'10 paper WEPD063 pages 3242-3244 Suppression schemem of COD (closed orbit distortion) caused by switching ripple in J-PARC 3GeV dipole magnet power supply
2. B.K. Kang et al., NIM A 417 (1998) 450-456 Reduction of ripple fields in a DC magnet using an auxilliary winding
3. shorted copper turns, H and V, of aspect ratio different from that (H) in figure 1, were suggested by M. Tiefenback. The first author's mis-understanding of the suggestion resulted in the configuration in figure 1.
4. http://www.sorensen.com/products/SG/SG_Specifications.htm
5. check suggested by Vladimir Kashikhin, FNAL

Acknowledgments

This work was improved by suggestions of G. Neil and A. Freyberger. We would like to thank M. Tiefenback for many stimulating discussions and helpful remarks, including finding references (1) and (2). This work was supported by by Jefferson Science Associates, LLC under U.S. DOE Contract No. DE-AC05-06OR23177. The U.S. Government retains a non-exclusive, paid-up, irrevocable, world-wide license to publish or reproduce this manuscript for U.S. Government purposes.